\documentstyle[prl,aps,twocolumn,epsf]{revtex}
\begin{document}

\twocolumn[\hsize\textwidth\columnwidth\hsize\csname
@twocolumnfalse\endcsname

\title{Bulk charge distributions on integer and fractional quantum Hall
plateaus}

\author{J. J. Palacios\cite{byline} and A. H. MacDonald}
\address{Department of Physics, Indiana University,
Bloomington IN 47405, USA}
\preprint{IUCM97-028}

\maketitle

\widetext
\begin{abstract}

\leftskip 2cm
\rightskip 2cm

We discuss the charge distributions across the bulk of a two-dimensional
electron gas system which is on an integer or fractional quantum Hall
plateau.  Our analysis is based on a relation, derived from the long
wavelength limit of the bulk density-density response function, between
the induced charge and leading derivatives of a slowly varying Hall 
potential.  We use the Wiener-Hopf method to solve the coupled response
and Poisson equations.  Unlike the integer case, treated in previous work, 
the induced charge in the fractional case can alternate in sign.

\end{abstract} 

\vskip2pc]

\narrowtext

\section{Introduction}

This paper addresses the change in the charge distribution and the
varying Hall field in the interior of a two-dimensional electron gas
which accompanies the dissipationless current flow of the integer and
fractional quantum Hall effects.  This quantity is of
interest,\cite{allan,experiments} in part, because the simple picture of
a constant Hall electric field generated by charges localized at the
surface of the sample, familiar from the three dimensional case, cannot
be replicated with two-dimensional electrostatics.  In addition, given
the quantized Hall {\it conductivity}, a non-constant Hall potential
implies a transport current in the bulk of the sample, which is
sometimes thought to be in conflict with the edge-state
picture\cite{edgestate} of the quantized Hall {\it conductance}.  The
induced charge in the bulk can, in principle, be measured, although
accurate measurements on the length scales of interest have proved
difficult\cite{experiments}.  In addition, unintentional and
uncharacterized disorder effects, not accounted for in the present work,
usually play a role in measurements on real samples.  

Our view, on which the work presented here is explicitly dependent, is
close to that carefully articulated by Thouless and coworkers in a
recent series of papers.\cite{thouless}  In particular, there {\it is}
generically a bulk contribution to the transport current under quantum
Hall conditions.  In typical macroscopic Hall bar samples, bulk current
tends to dominate\cite{endeffectcaveat} the total current flow.  This
view is {\it not} in contradiction with the edge state picture of the
quantized Hall conductance.  The paper is organized as follows.  In
Section II we derive a relationship between the charge density in the
bulk of a two-dimensional (2D) sample carrying a quantized Hall current
and leading derivatives of the Hall potential, presumed to be slowly
varying on microscopic length scales.   These equations are a
generalization of those first derived for the integer quantum Hall case
in Ref.  \onlinecite{allan}, and permit differences between integer and
fractional cases to be discussed\cite{kohmoto}.  The use of the
Wiener-Hopf method, first employed for this problem by
Thouless,\cite{thoulessold} is discussed in Section III.
We use it to express the solution to our coupled
response and Poisson equations in terms of a function
determined by numerical quadrature.
The bulk induced charge distribution depends in part on
non-universal aspects of the microscopic physics\cite{edgerefs} at the
edge of the sample where the bulk response relations do not apply.  In
the fractional case this leads to the appearance of a free parameter in
the solution which results from the Wiener-Hopf analysis.  In Section IV
we discuss classes of physically sensible solutions and contrast
behavior in the integer and fractional quantum Hall cases.  We conclude
in Section V with a brief summary of the paper. 

\section{Bulk response to a slowly-varying potential} 

The quantum Hall effect occurs\cite{leshouches} in a two-dimensional electron
system when, at zero temperature, 
there is a discontinuity in the chemical potential at an
electron density which depends on 
magnetic field strength.  At the point of the chemical
potential jump, the incompressible electronic ground state 
generally\cite{brokensymcaveat} has no gapless 
excitations.  In this section we discuss the response of electrons in
such an incompressible fractional Hall state to 
a slowly varying static electrostatic potential.  Bulk
response functions are usually described in terms
of their Fourier transforms:
\begin{equation}
n (q) = \Pi (q) \phi (q), 
\label{eq:polarization} 
\end{equation} 
where $\phi (q)$ includes both the external potential and 
the potential due to induced charges.  In a 2D system
the static density-density response function is related to
$\Pi (q)$ by
\begin{equation}
\chi (q) = \frac{\Pi (q)}{ 1 - V(q) \Pi (q)},
\label{eq:densdens}
\end{equation} 
where $V(q) = 2 \pi e^2/q$ is the 2D Coulomb interaction.

The precise form we use for $\Pi (q)$ below is based on 
a generalized single mode approximation\cite{oji,smalong}
for $\chi (q)$ in which the density operator is partitioned
into contributions associated with 
transitions between each pair of Landau levels.  In this approximation,  
which is expected to be reasonably accurate whenever the ground state is
incompressible, we obtain to leading order\cite{oji} 
in $(\hbar \omega_c)^{-1}$ at Landau level
filling factors $\nu \le 1$ 
\begin{equation}
\chi(q)  = \frac{-\nu}{\pi \ell^2}   \sum_{n=0}^{\infty}
\frac{s_n(q)}{\Delta_n(q)}.
\label{chiofq}
\end{equation}
Here $\Delta_n(q)$ is the collective mode energy for excitations
to the $n$-th Landau level,
\begin{equation}
s_n(q) = N^{-1} \langle \Psi_0 | \rho_{-q}^{0,n} \rho_q^{n,0} | \Psi_0 \rangle, 
\label{snofk}
\end{equation}
,$\omega_c = e B / m^* c$ is the cyclotron frequency,
,$\ell = (\hbar c / e B)^{1/2}$ is the magnetic length,
,$\nu = (2 \pi \ell^2) n$ is the density in units of the 
density which can be accomodated by a single Landau level,
$\rho_q^{n',n}$ is the contribution to the density operator 
associated with transitions between Landau levels $n$ and $n'$,
and $|\Psi_0\rangle$ is the many electron ground state 
when interaction terms which mix Landau levels are neglected.
It follows from Ref. \onlinecite{oji} that for $n \ne 0$
\begin{equation}
s_n(q) = \frac{ \exp ( -|q|^2 \ell^2 /2 ) } {n!} (q^2 \ell^2/2)^{n},
\label{snnot0}
\end{equation}
independent of the correlations in $|\Psi_0 \rangle$.
$s_0(q)$, on the other hand, does depend in detail on these
correlations although,\cite{smalong}
whenever $|\Psi_0\rangle$ is an incompressible state,
$s_0(q) \sim q^4$ at long wavelengths; the term proportional to 
$q^2$ in the long-wavelength expansion vanishes.  This property 
is responsible for the difference we discuss below between 
charge distributions across the interior for integer and fractional
incompressible states.

Keeping terms in up to $q^4$ in a long-wavelength expansion and 
up to $(\hbar \omega_c)^{-1}$ in a strong-field expansion we obtain
the following results for the electron number
density induced in an incompressible
bulk by a slowly-varying electrostatic potential:
\begin{equation}
n(x)= \frac{ \nu }{2 \pi \hbar \omega_c}
\frac{\partial^2 \phi(x)}{\partial x^2}
-\frac{C\ell^2}{e^2/\epsilon \ell} \frac{\partial^4 \phi(x)}{\partial x^4}
\;\;\;\;\;  
\label{induceddensity}
\end{equation}
where $\epsilon$ is the dielectric constant of the host semiconductor.
This result is obtained by noting that to the stated order in 
the strong field and long wavelength expansions  
$\Pi (q) = \chi (q)$ and by making the replacement valid for 
slowly varying potentials, $q \to -i \partial/ \partial x$.  
The first term on the right hand side will always dominate for 
sufficiently slowly varying densities and was derived  
originally\cite{allan} via a different line of argument valid 
only for the special case of non-interacting electrons and 
integer filling factors.  The present derivation has generic validity.
The second term is retained\cite{caveatq4}
here because of its larger relative importance at stronger 
magnetic fields.  The coefficient $C$ in this term vanishes in the case of 
the incompressible states which occur at integer Landau level
filling factors and is not accurately known for all fractional
incompressible states.  In the case of $\nu = 1/3$ and $\nu = 2/3$ 
it follows from Ref. \onlinecite{smalong} that 
\begin{equation}
C = { e^2 / \epsilon \ell}{12 \pi \Delta (q \to 0)} \sim 0.2.
\label{cvalue}
\end{equation}
For other incompressible states with smaller gaps, its value 
is likely to be larger.

Comparing the two terms motivates 
the introduction of the length scale $L$ on which the two
terms have comparable importance:
\begin{equation}
\frac{L}{\ell} = \big[ \frac{2 \pi C}{\nu}\frac{\hbar \omega_c} 
{e^2/ \epsilon \ell} \big]^{1/2}
\label{lengthscale}
\end{equation} 
This length scale diverges in the limit of strong fields
but is not extremely long in typical experimental 
situations; for example for $\nu = 1/3$ and $B = 
20$ Tesla, $L \sim 3 \ell$.   The fact that the induced
charge density is proportional to derivatives of the 
electrostatic potential rather than to the potential
itself is a direct consequence of the incompressibility of the 
many-electron ground state.  For compressible states, Thomas-Fermi
screening theory gives 
\begin{equation}
n(x) = - (d n / d \mu) \phi (x)
\label{thomasfermi}
\end{equation} 
where $dn/ d\mu$ is the thermodynamic density of states of the 
uniform bulk state.  This quantity is zero at zero temperature 
for incompressible states.  
However, it will\cite{caveatdisorder} be finite at 
any finite temperature.  We can define a 
length scale $L'$ at which the Thomas-Fermi screening 
term in Eq. ~\ref{thomasfermi} and the leading derivative 
term in Eq. ~\ref{induceddensity} become comparable:
\begin{equation}
\frac{L'}{\ell} \sim  \big[ \frac{k_B T}{\Delta } \big]^{1/2} 
\exp (\Delta  / 4 k_B T)
\label{thomasfermilength}
\end{equation} 
where $\Delta$ is the chemical potential
jump at the density of the incompressible state.
Our neglect of the Thomas-Fermi screening term implicitly
assumes $L'$ is longer than any length of interest
so that we require that $k_B T \ll \Delta/ 4$.

\section{Wiener-Hopf method}

Eq. ~\ref{induceddensity}  can be rewritten using $\ell$ 
as the length unit:
\begin{equation}
\tilde \rho (\tilde x)=\alpha \frac{\partial^2 \phi(\tilde
x)}{\partial\tilde x^2} 
-\beta \frac{\partial^4 \phi(\tilde x)}{\partial \tilde x^4},
\label{density}
\end{equation}
where $ \tilde \rho(\tilde x)= \ell^2 n (\tilde x)$,
$\alpha= \nu /(2 \pi \hbar \omega_c)$, $\beta= C/(e^2/\epsilon \ell) 
= \alpha L^2 / \ell^2 $, and $\tilde x =  x/\ell$. 
We will drop the tildes used to distinguish dimensionless
quantities in Eq. ~\ref{density}  in what follows.

We choose a coordinate system where all points with $x > 0 $ 
are sufficiently deep in the bulk that Eq. ~\ref{density}  applies.
The electrostatic potential appearing here 
is given by the Poisson equation:
\begin{equation}
\phi(x)= -\lambda  \int_0^\infty dx' \; \rho(x') \log{(|x - x'|)} 
+ \phi_{ext}(x)
\label{pot},
\end{equation}
where $\lambda=2 e^2 /\epsilon \ell $.  Here $\phi_{ext}(x)$ specifies
the potential created by the electronic charge density for $x < 0$ and by 
charges external to the electron system.  We will initially drop this 
contribution to Eq. ~\ref{pot} ; in the end it will implicitly be
accounted for in the non-universal boundary conditions
that must be imposed in order to fix the charge distribution.

We start by Fourier transforming equation Eq.\ \ref{pot} : 
\begin{eqnarray}
\phi_-(k) + \phi_+(k) & = &- \lambda v(k) \rho_+(k)
\label{ft},
\end{eqnarray}
where we define the Fourier transforms as follows:
\begin{eqnarray}
v(k)&=&\int_{-\infty}^\infty dx \;e^{ikx} \log (|x|) \\
\rho_+(k)&=&\int_0^\infty dx\; e^{ikx}  \rho(x) \\
\phi_+(k)&=&\int_0^\infty dx \; e^{ikx} \phi(x) \\
\phi_-(k)& =&\int_{-\infty}^0 dx \; e^{ikx} \phi(x).
\label{ftdef}
\end{eqnarray}
Application of the Wiener-Hopf method\cite{wienerhopf} requires  
the continuation of the Fourier 
transform wavevector to the complex plane: 
$k=\sigma + i \tau$.  To ensure convergence of spatial integrals
a convergence factor, $\exp (-a |x|)$, is included in the 
interaction kernel, so that $v(k)$ is regular and bounded 
for $|\tau| < a$.  $v(k)$ 
then has branch cuts along the portions of the imaginary axis with 
$|\tau| > a$.  (For completeness, explicit expressions 
for $v(k)$ are given in the Appendix.  No difficulties are 
encountered in setting $a \to 0$ at the end of the calculation.)
The $+$ and $-$ subscripts in the 
above definitions are used to designate functions which 
are regular and bounded in the upper and lower halves 
of the complex $k$-plane, respectively.
Eq. ~\ref{ft}  contains three unknown functions. 
Using the response relation for $x > 0$ and integrating the half
Fourier transform by parts we obtain 
\begin{eqnarray}
\rho_+(k)& =& \beta \phi'''(0) -i\beta k\phi''(0) - (\alpha+\beta
k^2)\phi'(0) + \nonumber \\
&&(i\alpha  k+ i\beta k^3) \phi(0) -(\alpha k^2 + \beta k^4) \phi_+(k)
\label{rhopk}
\end{eqnarray}
which can be solved for $\phi_{+} (k)$:
\begin{eqnarray}
\phi_+(k)&=&\frac{1}{\alpha k^2 + \beta k^4} [\beta \phi'''(0) -i\beta
k\phi''(0) - (\alpha+\beta k^2)\phi'(0) + \nonumber \\
&& (i\alpha  k+ i\beta k^3)\phi(0) - \rho_+(k)].
\label{phipk}
\end{eqnarray}
Here we have chosen $\phi (\infty) = 0$ and assumed that 
all derivatives of $\phi (x)$ vanish at $\infty$.  
Note that if we had retained the external potential term, 
the form of this equation would not have changed.  
Combining this with Eq. ~\ref{ft} yields 
\begin{equation}
A(k)\phi_-(k)+B(k)=\rho_+(k) K(k),
\label{w-h}
\end{equation}
where
\begin{eqnarray}
A(k)&=&\alpha k^2 + \beta k^4, \nonumber \\
B(k)&=&\beta \phi'''(0) -i\beta k\phi''(0)  -  (\alpha+\beta k^2)\phi'(0)
+ (i \alpha k + i \beta k^3)\phi(0)   \nonumber \\
K(k)&=&1-\lambda A(k) v(k).
\end{eqnarray}

The essential difficulty 
in the application of the Wiener-Hopf method occurs in making the following 
factorization of $K(k)$: 
\begin{eqnarray}
K(k)= K_+(k)/K_-(k),
\end{eqnarray}
where $K_+(k)$ is regular and bounded in the upper half
plane, $K_-(k)$ is regular and bounded in the lower half
plane, and both functions share a common strip of regularity along
the real line.  In some cases this factorization, which is not
in general unique, can be established by inspection and $K_{\pm}(k)$ 
specified analytically.  Here we resort to a general method 
with wide applicability\cite{wienerhopf}.  The  
factors will be specified by a function determined by 
a numerical integration.
The method starts by defining $ q(k) = \ln [K (k)] $.
Since $K(k)$ has no zeroes, at least near the real line,
it follows from the Cauchy integral theorem 
that, for any positive $\xi < a$, \,   
$q(k) = q_+(k) - q_-(k)$ where 
\begin{equation}
q_{\pm}(k)= \frac{1}{2\pi i} \int_{-\infty }^{\infty} d \chi
\frac{q(\chi \mp i \xi) }{\chi \mp i \xi - k}.
\label{qofk}
\end{equation} 
By construction, $q_{+}(k)$ is regular and bounded in the upper 
half plane and $q_{-}(k)$ in the lower half plane.  
This device provides the required factorization since
$K(k) =  K_+(k)/K_-(k)$ where $K_{\pm}(k) = \exp [q_{\pm}(k)]$.

\begin{figure}
\epsfxsize=8cm \epsfysize=8cm \epsfbox{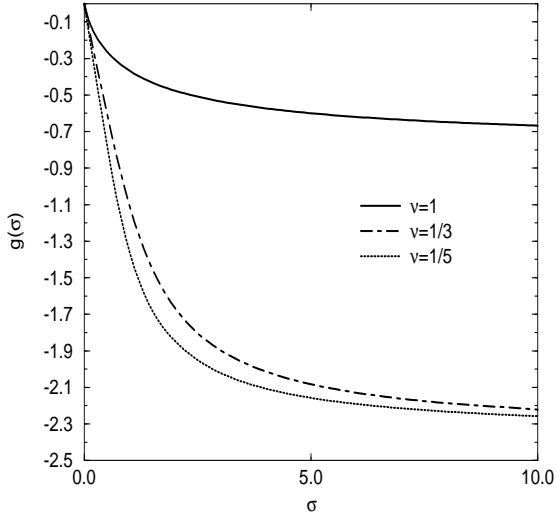}
\caption{Numerically calculated $g(\sigma)$ for
$\nu=1$ ($ L/\ell=0$ ), 1/3 ($ L/\ell=1.7$), and 1/5 $(L/\ell=3)$ at
$B=10$T.  The two fractional cases give similar results.}
\label{fig1}
\end{figure}

Given the required factors we can reexpress Eq.\ \ref{w-h} as
\begin{equation}
A(k)\phi_-(k)K_-(k)+B(k)K_-(k)= \rho_+(k)K_+(k)
\label{neww-h}
\end{equation}
The left and right hand sides of Eq. ~\ref{neww-h}  are
regular and bounded in the lower and upper planes, respectively, and they
share a strip of regularity of width $2 a$ along the real axis. 
It follows from analytic continuation that both sides must be 
equal to an as yet unknown polynomial $P(k)$, which is regular 
in the whole complex plane. 
To evaluate the induced charge density in the 
bulk we require $\rho_+(k)$, and hence $K_+(k)$, only along the real axis.
For real $k$, choosing $\xi$ to be infinitesimal in Eq. ~\ref{qofk} gives  
\begin{equation}
q_{\pm}(\sigma) = \pm \frac{1}{2} q(\sigma) + i g(\sigma)
\label{betainf}
\end{equation}
where
\begin{equation}
g(\sigma)=-\frac{1}{2\pi}
{\rm P} \int_{-\infty}^{\infty} \frac{\log K(\chi)}{\chi - \sigma} \;d\chi.
\label{gofsigma}
\end{equation} 
In this final expression we can remove the convergence 
factor in the interaction kernel by setting $a \to 0$. 
We evaluate the integral numerically; the weak divergences for 
$\chi \to \pm \infty$ cancel and create no difficulty.  Exponentiating
gives the required factors: $K_{\pm}(\sigma) =
[K(\sigma)]^{\pm 1/2} \exp [ i g (\sigma)]$.
Figure \ref{fig1} shows $g(\sigma)$ for $\nu=1$ 
($L/\ell=0$), $\nu=1/3$ ($L/\ell=1.7$), and $\nu=1/5$ ($L/\ell=3$)
at $B=10$ in the three cases. For
the integer case we recover $g(\sigma \to \infty ) = - \pi/4$,
in agreement with the analytic factorization\cite{thouless} while
for the fractional case $g(\sigma \to \infty)  = - 3  \pi /4$; this 
difference can be traced to the fact that at large 
$\sigma$, $K (\sigma) \sim \sigma$ 
in the integer case and $\sim \sigma^3$ in the fractional case.

The Wiener-Hopf factorization leads to the following equation which 
is used below to calculate the induced charge density: 
\begin{eqnarray}
\rho_+(\sigma)& = &P(\sigma)/K_+(\sigma)\nonumber \\
& = & P(\sigma) 
\{1 + \nu  (e^2/\epsilon \ell)/ (\hbar \omega_c) |\sigma| [ 1 +
\sigma^2 (L/\ell)^2] \}^{-1/2} \times \nonumber \\
&&\exp[-i \; g(\sigma)],
\label{rhoofk}
\end{eqnarray}
The polynomial $P(\sigma)$ can be determined by analysis of the limits
$\sigma \rightarrow 0$ and $\sigma \rightarrow \infty$.  In the present
case the requirement that $\rho_+(\sigma)$ vanishes at large $|\sigma|$
implies that $P(\sigma)=P_0+P_1\sigma$ in the fractional case and
that $P(\sigma)=P_0$ in the integer case where $L/\ell=0$.  Taking
$\sigma \rightarrow 0$ in Eq.\ \ref{rhopk}, relates $P_0$ and $P_1$ to
$\phi(0)$ and its first three derivatives.   Note that $P_1$ is pure
imaginary as required in order to obtain $\rho_+(x)$ real. 

\section{Induced charge distributions}

\begin{figure}
\epsfxsize=8cm \epsfysize=8cm \epsfbox{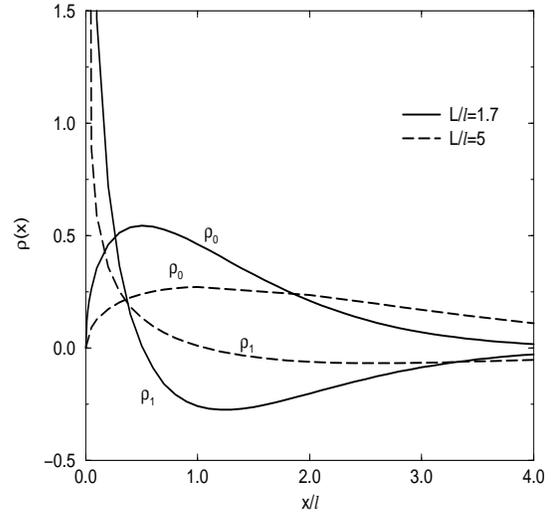}
\caption{The two contributions, $\rho_0$ and $\rho_1$, (see text)
to the total electronic density distribution for $\nu=1/3$ at $L/\ell=1.7$
and $L/\ell = 5$.  This results were calculated for $B=10$T.}
\label{fig3}
\end{figure}

Inverting the Fourier transform in Eq.\ \ref{rhoofk}
gives for $x > 0$ $\rho(x) = P_0 \rho_0(x) +  i P_1 \rho_1(x)$. 
These two contributions to the total charge are given by
\begin{eqnarray}
\rho_0(x) &=&\frac{1}{\pi}\int_0^\infty d\;\sigma \frac{
\cos [g(\sigma)+ \sigma x]}{ \{1 + \nu
(e^2/\epsilon \ell)/ (\hbar \omega_c) \sigma [ 1 +
\sigma^2 (L/\ell)^2] \}^{1/2}} \\
\rho_1(x) &=&\frac{-1}{\pi}\int_0^\infty d\;\sigma \frac{
\sigma \sin [g(\sigma)+ \sigma x]}{\{ 1 + \nu
(e^2/\epsilon \ell)/ (\hbar \omega_c) \sigma [ 1 +
\sigma^2 (L/\ell)^2] \}^{1/2}} 
\label{rhoarhob} 
\end{eqnarray}
Previous results\cite{thouless,thoulessold}
for integer plateaus are recovered by setting 
$P_1$ and $L$ to zero.  
Numerical results for $\rho_0(x)$ and $\rho_1(x)$ calculated for
$\nu=1/3$ at $B=10$T are shown in Fig.\ \ref{fig3} for two values of
$L/\ell$.  The fact that $\rho_0(0)=0$ follows from the 
analyticity of $K_{+}(\sigma)$ in the upper half-plane.  The 
square root behavior at small $x$, evident in Fig.~\ref{fig3},
follows from the fact that $g(\sigma)$ approaches a 
constant at large $\sigma$ and from the $\sigma^{+3/2}$ behavior 
of the denominator of the integrands in Eq.~\ref{rhoarhob}.
Because of the extra factor of $\sigma$ in the numerator
of its integrand,
\begin{equation}
\rho_1(x) \sim  - \frac{ \sin [g(\infty)]}  
{ x^{1/2} (L/\ell)[ (e^2/\epsilon \ell)/ (\hbar \omega_c)]^{1/2}} 
\end{equation}
for small $x$. 

\begin{figure}
\epsfxsize=8cm \epsfysize=8cm \epsfbox{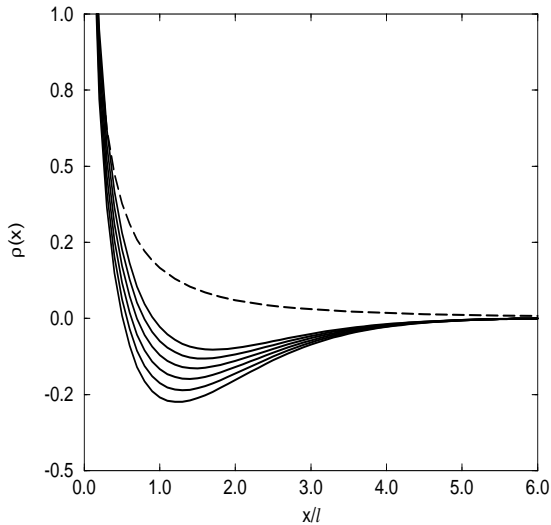}
\caption{Total electronic density distribution, $\rho(x)$ (normalized to
$iP_1$) 
at $\nu=1/3$ and $B = 10$T ($L/\ell=1.7$) for $P_0/iP_1=0.0,0.1,0.2,0.3$,
 and 0.5 (bottom to top). 
The charge distribution for the integer case (dashed line) is also
shown for comparison (normalized to the coefficient $P_0$). }
\label{fig4}
\end{figure}

In Fig. \ref{fig4} we plot $\rho(x)/iP_1$ for a series of values of
$P_0/iP_1$.  The ratio $P_0/iP_1$ depends on non-universal details of
the microscopic physics at the edge of the sample and cannot be
determined by the present analysis.   Notice that, unlike the integer
case, the charge distribution can change in sign as a function of
distance from the edge.  The induced charge density in integer case,
which also diverges as $x^{-1/2}$ for $x \to 0$, is also shown in
Fig.~\ref{fig4} (normalized to $P_0$ in this case).

\section{Summary} 

The work reported in this paper is based on a new derivation
of the relationship between the induced charge in the bulk
of a quantum Hall sample and derivatives of a slowly
varying Hall potential.  The derivation is based on 
gradient expansion of the long-wavelength bulk response 
functions.  It generalizes results obtained previously
for the case of the integer quantum Hall effect to 
the case of interacting electrons, and suggests that 
at fractional filling factors and especially at very
strong magnetic fields, a higher fourth derivative term 
must be included in the induced density expression.  
The approach taken here is more surely grounded by
microscopic physics than related recent work based on
Chern-Simons mean field theories,\cite{kohmoto} although
the physical conclusions appear to be quite similar.
We have solved the coupled Poisson and response equations
to determine the self-consistent charge density distribution
in the interior of a Hall bar on a quantized platueau.
Our solution is based on the Wiener-Hopf method 
in which the required factorization of the response function
is specified in terms of a function which is determined 
by numerical evaluation of an integral.  We find that, in 
the fractional case, the induced charge density  
can alternate in sign in the interior of the sample. 

\acknowledgements 

This work was supported by the National Science Foundation under
grant DMR-9714055.  The authors acknowledge helpful interactions
with M. Kohmoto, J. Shiraishi, D. J. Thouless, and C. Wexler. 

\section*{Appendix}

Including a convergence factor in the integral,
the Fourier transform of $v(x)$ at general complex wavevector
$k$ is given by: 
\begin{eqnarray}
v(k)&=&2\int_0^{\infty} dx\; \log(x) \cos[kx] e^{-ax}\nonumber \\
&=&\frac{i}{k}\log[(1+ik/a)/(1-ik/a)]
\end{eqnarray}
For $a \to 0$ and $k = \sigma + i \tau$ 
\begin{equation}
v(k) = -\frac{\pi|\sigma|}{\sigma^2 + \tau^2} + i \frac{\pi \tau}
{\sigma^2 + \tau^2} \;Sign(\sigma). 
\end{equation}
$\Re[v(k)]$ is continuous everywhere but $\Im[v(k)]$ is discontinuous
along the the entire imaginary axis.


\begin{references}
\bibitem{endeffectcaveat}  In making this assertion we have in mind a
classification based on integrating the current density across the Hall
bar far enough from the current source and drain to avoid end effects.
\bibitem{caveatq4} We have dropped terms proportional
to $q^4$ in $\Pi (q)$ which vanish for $(\hbar \omega_c)^{-1} \to 0$.
\bibitem{caveatdisorder} The plateau in the quantized Hall conductance 
in a disordered systems generally includes regimes in which 
localized charged excitations are present at the Fermi 
energy.  The thermodynamic density of states is then 
non-zero.  Present considerations apply only to weakly disordered 
systems and in particular require that the thermodynamic 
density of states at the mid-point of a quantum Hall plateau be small.
\end{references}
\end{document}